\documentclass[prc,twocolumn,superscriptaddress,showpacs,floatfix,aps,10pt]{revtex4-1}
\usepackage{graphicx,amssymb,nicefrac}
\usepackage{amsfonts}
\usepackage{times}
\usepackage{dcolumn}
\usepackage{bm}
\usepackage{braket}
\usepackage[usenames,dvipsnames,svgnames,table]{xcolor}
\usepackage{multirow} 
\usepackage{diagbox}
\usepackage{placeins}
\usepackage{amsmath}
\usepackage{url}
\usepackage[outdir=./]{epstopdf}
\renewcommand{\vec}[1]{\mbox{\boldmath $#1$}}

\newcommand{\simleq}{\; \raisebox{-0.4ex}{\tiny$\stackrel
{{\textstyle<}}{\sim}$}\;}

\usepackage{color}
\definecolor{darkblue}{rgb}{0.,0.,0.4}
\definecolor{darkred}{rgb}{0.5,0.,0.}
\usepackage[colorlinks=true,linkcolor=blue,citecolor=blue,urlcolor=blue]{hyperref}

\mathchardef\myhyphen="2D

\begin{document}

\preprint{ }

\title{Quantified Gamow Shell Model interaction for \mbox{\boldmath $psd$}-shell nuclei}

\author{Y. Jaganathen}
\affiliation{NSCL/FRIB Laboratory, Michigan State University, East Lansing, Michigan 48824, USA}

\author{R. M. Id Betan}
\affiliation{Physics Institute of Rosario (CONICET), Rosario, Argentina}
\affiliation{Department of Physics and Chemistry FCEIA (UNR), Rosario, Argentina} 
             
\author{N. Michel}
\affiliation{NSCL/FRIB Laboratory, Michigan State University, East Lansing, Michigan 48824, USA}

\author{W. Nazarewicz}
\affiliation{Department of Physics and Astronomy and NSCL/FRIB Laboratory, Michigan State University, East Lansing, Michigan 48824, USA}

\author{M. P\l{}oszajczak}
\affiliation{Grand Acc\'el\'erateur National d'Ions Lourds (GANIL), CEA/DSM - CNRS/IN2P3, Caen, France}

\date{\today}

\begin{abstract} 
  \begin{description} 
  
    \item[Background] The structure of weakly bound and unbound nuclei close to particle drip lines is one of the major science drivers of nuclear physics.  A comprehensive understanding of these systems goes beyond the traditional configuration interactions approach formulated in the Hilbert space of localized states (nuclear shell model) and requires an open quantum system description. The complex-energy  Gamow Shell Model (GSM) provides such a framework as it is capable of describing  resonant and non-resonant many-body states on equal footing.

    \item[Purpose] To make reliable  predictions, quality  input is needed that  allows  for the full uncertainty quantification of theoretical results.  
    In this study, we carry out the optimization of an effective GSM (one-body and two-body) interaction in the  $psdf$ shell model space. The resulting interaction is expected to describe  nuclei with $5 \leq A \simleq 12$ at the   $p-sd$-shell interface.

     \item[Method]  The one-body potential of  the $^4$He core is modeled by a Woods-Saxon + spin-orbit + Coulomb potential, and the  finite-range nucleon-nucleon interaction between the valence nucleons consists of central, spin-orbit, tensor, and Coulomb terms. The GSM is used to compute key fit-observables. The chi-square optimization 
     is performed using the Gauss-Newton algorithm augmented by  the singular value decomposition technique. The resulting covariance matrix enables quantification of statistical errors within the linear regression approach.
          
    \item[Results] The optimized one-body potential reproduces  nucleon-$^4$He scattering phase shifts up to an excitation energy of 20 MeV. The two-body interaction built on top of the optimized one-body field
     is  adjusted to the bound and unbound ground-state binding energies   and selected excited states of the  Helium, Lithium, and Beryllium isotopes up to $A=9$. A very good agreement with experiment was obtained 
    for binding energies.  First applications of the optimized interaction include predictions for two-nucleon correlation densities and excitation spectra of light nuclei with quantified uncertainties. 
 
    \item[Conclusion] The new interaction will enable comprehensive and fully quantified studies of structure and reactions aspects of nuclei from the $psd$ region of the nuclear chart. 
    
\end{description}
\end{abstract}

\maketitle

\section{Introduction}\label{sec.introduction}

Light nuclei have traditionally provided  an excellent testing ground for microscopic nuclear structure models. The pioneering work on $p$-shell nuclei by Lane and Kurath \cite{1955Lane,1956Kurath}  set the foundations of the interacting Shell Model (SM) \cite{2005Caurier}, which became the cornerstone of nuclear structure theory and provided the guidance for understanding the wealth of spectroscopic data on energy levels, electromagnetic transitions, nuclear moments, and various particle decays. The progress in SM description of nuclei has been  achieved thanks to the identification of basic features and symmetries of the bare nucleon-nucleon interaction, and the continuous efforts to develop reliable effective interactions in different valence subspaces. Nowadays, with the progress in radioactive beam experimentation and the development of 
microscopic $A$-body nuclear models 
(see Refs.~\cite{2013Forssen,Hagen14,Lahde14,Carlson15,Hergert2016,Navratil16} for recent reviews), light nuclei still  remain the favorite laboratory for testing both the nuclear interactions and  many-body approaches.

Nuclei in the vicinity of drip lines provide great challenge for the nuclear theory due to the key role of coupling to the scattering continuum and the decay channels \cite{Transactions,Dob07,GSMReview}. The challenge for theory is to develop methodologies to reliably calculate and understand properties of new physical systems with  large neutron-to-proton asymmetries and low-energy reaction thresholds. 
The impact of resonances and the non-resonant scattering continuum on nuclear properties can be only considered in the open quantum system formulation of the SM (see Ref. \cite{2003Oko} and references quoted therein).  An approach that provides the rigorous treatment of the many-body correlations and the coupling to the resonant and non-resonant particle continuum, is the complex-energy continuum SM  based on the Berggren ensemble, the Gamow Shell Model (GSM) \cite{GSMReview,2002MichelPRL,2003MichelPRC,2004MichelPRC,2002Betan,2003Betan}. GSM can be considered an open quantum system extension of the interacting SM. It can be formulated equivalently in the Slater determinant representation or in the reaction channel representation. In the latter case GSM can also describe reaction cross sections \cite{2014Jaganathen,2015Fossez,2017Dong}. 

A profound  challenge for modern nuclear theory is to be able to compute   uncertainties of theoretical
predictions. Indeed, without uncertainty quantification of theoretical results, it is impossible to
assess the quality of predictions, especially when it comes to  extrapolations, and discriminate between different models (see Refs.~\cite{Editors2011,2014Dobaczewski,Naz16}). Therefore, evaluating errors on predicted quantities is nowadays an  essential component of nuclear modeling. 
In the context of nucleon-nucleon potentials and few-body systems, the systematic work on uncertainty quantification has just begun  \cite{Ama13,2014NavarroPerez,2014NavarroPerez_Bootstrapping,Piarulli15,Piarulli16,2013Ekstrom,Eks15,Car16}.

Our aim is to provide a  practical approach to light nuclei by separating the model space into the core and valence shells and determining the effective one-body core-nucleon potential and the finite range two-body  interaction between valence nucleons that consists of central, spin-orbit, tensor, and Coulomb terms. The interaction -- constrained by key binding energies  of He, Li, and Be isotopes with $A\leq 9$ -- is 
optimized in a large $psdf$   space involving resonant and scattering states, with the overall  goal of describing structure and reaction aspects of light  nuclei with $A \simleq 12$. 
To provide quantified predictions, we  carry out statistical analysis of our model.
To this end, we use the  singular value decomposition technique  to 
estimate the effective size of the model parameter space, or  indeterminacy of parameters.
The statistical uncertainties on the parameters and observables  are then evaluated by  using the covariance matrix obtained within the linear regression approach. The resulting  GSM interaction is then applied to predict two-nucleon correlation densities and excitation spectra of selected light nuclei with quantified uncertainties.

This paper is organized as follows. In Sec.~\ref{Sec.Framework}, we review the basics of GSM calculation (Sec. \ref{SSec.GSM}) and describe the  form of the effective interaction (Sec. \ref{SSec.GSMint}) used in this work. Section~\ref{Sec.Optimization} outlines the  optimization methodology applied  and the framework  of   uncertainty quantification adopted. The optimization of the core-nucleon interaction is presented in Sec.~\ref{Sec.WS} while Sec.~\ref{Sec.EffectiveInteraction} deals with the optimization of valence-nucleons interaction  and contains  results for ground state energies of He, Li, and Be with $A\leq 9$. Several applications of the optimized GSM interaction are given in 
Sec.~\ref{Sec.Applications}. In particular, we discuss the two-nucleon correlation densities in $^6$He and $^6$Li (Sec. \ref{SSec.Correlation_densities}) and present the quantified predictions for excited  states in selected isotopes of He, Li, and Be (Sec. \ref{SSec.Excited_states}). Finally, the main conclusions of this work are summarized in Sec.~\ref{Sec.conclusions}.

\section{The framework} \label{Sec.Framework}

\subsection{The Gamow Shell Model} \label{SSec.GSM}

The Gamow Shell Model \cite{GSMReview,2002MichelPRL,2003MichelPRC,2004MichelPRC,2002Betan,2003Betan} is an open-quantum system extension of the traditional shell model formulated in the complex momentum $k$-plane.  In the GSM, the single-particle (s.p.) basis is the Berggren basis \cite{Berggren1,Berggren2} generated by a finite-depth potential. The Berggren ensemble  consists of Gamow (or resonant) discrete states and the non-resonant scattering continuum; it satisfies the completeness relation:
\begin{equation}\label{Eq.Berggren}
  \sum_{n \in \{b,d\}} \vert u_{n\ell j} \rangle \langle \tilde{u}_{n\ell j} \vert + \int_{\mathcal{L}^+_{\ell j}} \!\!\vert u_{k\ell j}  \rangle \langle \tilde{u}_{k\ell j} \vert d k = 1
\end{equation}
that holds for each $(\ell,j)$ partial wave. $\vert u_{n \ell j} \rangle$ are the radial wave functions of the bound (b) and decaying (d) states, $\vert u_{k \ell j}\rangle$ are scattering states, and $\mathcal{L}^+_{\ell j}$ is a contour that encompasses the decaying states in the fourth-quadrant of the complex $k$-plane. This basis can describe any s.p. bound state, as well as any s.p. decaying resonance provided its complex momentum $k$ is located between the contour $\mathcal{L}^+_{\ell j}$
and the real axis \cite{Berggren1,Berggren2}. While bound states satisfy the standard normalization,  the decaying states are normalized using the complex scaling method \cite{1971Gyarmati} and the scattering states are normalized to Dirac delta distributions. In practice, the contours $\mathcal{L}^+_{\ell j}$ are discretized using a Gauss-Legendre quadrature with a minimum of 30 points to ensure an acceptable degree of  completeness. The momentum cutoff in the integral has to be sufficiently large to ensure convergence; for this study, the value  $k_{\mbox{\scriptsize{max}}} = 2$~fm$^{-1}$ was chosen.

The Slater determinants spanned by the s.p. Berggren basis states define the many-body basis in which the Hamiltonian is diagonalized. Assuming that the nucleus can be described as a system of $N_v$ valence nucleons outside the closed core, the GSM Hamiltonian, expressed in the intrinsic nucleon-core  Cluster Orbital Shell Model (COSM) coordinates \cite{Suzuki1988}, can be written as:
\begin{equation}
H = \sum_{i=1}^{N_v} \left[ \frac{\vec{p}_i ^2}{2\mu_i} + U_{\mbox{\scriptsize{core}}}(i) \right] +\!  \sum_{i<j=1}^{N_v} \!\left[ V(i,j) + \!\frac{\vec{p}_i \vec{p}_j}{M_{\mbox{\scriptsize{core}}}}  \right], \label{Hamiltonian}
\end{equation}
where $U_{\mbox{\scriptsize{core}}}$ is the s.p. core-nucleon potential, $V$ is the two-body interaction between the valence nucleons, and  $\mu$ and $M_{\mbox{\scriptsize{core}}}$ stand respectively for the reduced mass of the nucleon and the mass of the core. The last term in Eq.~(\ref{Hamiltonian}) represents the two-body recoil term. The GSM-COSM Hamiltonian is by construction translationally invariant, and the error on binding energies that arises from the approximate coordinate antisymmetry in the center-of-mass frame is shown to be smaller than 2\% in comparison to a fully-consistent calculation in the Jacobi coordinates \cite{2017Simin}.

To optimize calculations, the s.p. basis  is adapted for each nucleus. First, a Hartree-Fock (HF) procedure is carried out with the Hamiltonian (\ref{Hamiltonian}), and gives the complex energies of the  optimized $0p_{3/2}$ and $0p_{1/2}$ poles. For the very light nuclei, these poles can become very unbound and result in imprecise s.p. wave functions. To remedy this problem, the HF potentials are replaced by Woods-Saxon (WS) potentials with the same complex-energy  $0p_{3/2}$ and $0p_{1/2}$ poles. It is worth noting that thanks to the inclusion of the Coulomb potential directly in the basis, the outgoing s.p. proton wave functions have the correct  asymptotic behaviors at infinity. The diagonalization of the Hamiltonian in the $N_v$-body Berggren Slater determinant basis generated by the basis potential  provides the many-body bound and resonant states. All details pertaining to the solution of the GSM eigenproblem  can be found in the review \cite{GSMReview}.

The main challenge in GSM applications comes with the presence of  matrices of huge rank that arise from the discretization of the continuum contours $\mathcal{L}^+_{\ell j}$. As a result, GSM calculations are usually performed in truncated spaces in which only a relatively few  particles are allowed in the scattering continuum. Many-body techniques such as the Density Matrix Renormalization Group (DMRG) \cite{DMRG1,DMRG2} offer ways to overcome this dimension barrier by including the continuum couplings progressively. In this article, we make use of the natural orbitals \cite{NaturalOrbitals} defined as the eigenvectors of the (approximate) one-body density matrix $\rho'_{mn} = \langle \Psi'  \vert \left[ a^\dagger _m \tilde{a}_n \right]^0_0\vert  \Psi' \rangle $ where $m$ and $n$ are the Berggren basis states, and $\vert \Psi' \rangle$ an approximation of the final many-body state. Compared to the  Hartree-Fock states, natural orbitals offer a more adapted basis to the diagonalization problem, on the assumption that $\vert \Psi' \rangle$ is close to the desired final many-body state. This technique has been applied with  success in the contexts of Variational Multiparticle-Multihole Configuration Mixing Method \cite{NaturalOrbitals0} and DMRG \cite{2017Shin}. In the present study, the approximate solution $\vert \Psi' \rangle$ is obtained in a smaller configuration space in which only two particles are allowed in the non-resonant continuum space. With the corresponding natural orbital basis, a number of $5-7$ states per partial wave offer results of similar quality ($\Delta E<15$ keV) as the original 30 Berggren basis states; this  sometimes reduces the sizes of the matrices by four orders of magnitude, thus making  large-space calculations  tractable.

\subsection{The GSM interaction}\label{SSec.GSMint}

In this study, the light nuclei are described in terms of valence nucleons  outside the $^4$He core. As seen in Eq.~(\ref{Hamiltonian}), the GSM interaction has two components: the one-body core-valence potential $U_{\mbox{\scriptsize{core}}}$  and the two-body interaction $V$ between the valence nucleons. The core-valence potential is modeled, separately for protons and neutrons, by the sum of a Woods-Saxon potential, a spin-orbit term, and a Coulomb field: 
\begin{equation}
  U_{\mbox{\scriptsize{core}}}(r) = V_0 f(r) - 4V_{\ell\;\!\!s} \frac{1}{r} \frac{df(r)}{dr}\, \vec{\ell} \cdot \vec{s} + U_{\mbox{\scriptsize{Coul}}}(r) \: \label{eq.woodsSaxon}
\end{equation}
where $f(r)=-(1+\exp[(r-R_0)/a])^{-1}$. The  WS potential depth $V_0$, the spin-orbit strength $V_{\ell\;\!\!s}$, the radius $R_0$ and the diffuseness $a$  are the four parameters that enter the  optimization carried out independently for  protons and neutrons. The Coulomb potential for protons $U_{\mbox{\scriptsize{Coul}}}$ was kept fixed and equal to the potential generated by a spherical Gaussian charge distribution. It can be cast in the form $ U_{\mbox{\scriptsize{Coul}}}(r) = 2e^2\: \mbox{erf}(r/\tilde{R}_{\rm ch}) / r$ \cite{1977Saito}, where
$\tilde{R}_{\rm ch}= 4R_{\rm ch}/(3\sqrt{\pi})$ and
the experimental value of  charge radius of $^4$He is $R_{\rm ch}=1.681$\,fm \cite{Sick}.  

A general form of a two-body effective nuclear potential was derived in the early 1940s \cite{1958Okubo,1941Eisenbud}, when a tensor potential was added in addition to central and two-body spin-orbit potentials to describe the quadrupole moment of the deuteron. The first applications of such an interaction using Gaussian form factors  succeeded in  reproducing nucleon-nucleon (NN) scattering data up to 300 MeV \cite{1957Gammel}. In this paper, we shall use a NN-potential which is  a sum of central, spin-orbit, tensor, and Coulomb terms:
\begin{equation}
V = V_c + V_{\scriptscriptstyle{L\:\!\!S}} +V_{\scriptscriptstyle{T}} +V_{\mbox{\scriptsize{Coul}}} .
\end{equation}
The two-body Coulomb potential $V_{\mbox{\scriptsize{Coul}}}(r) = e^2/r$ between  valence protons is treated exactly by incorporating its long-range part into the basis potential  (see Ref. \cite{2010Michel} for a detailed description of the method). The central, spin-orbit and tensor part of the interaction are based on an interaction introduced in Ref. \cite{1979Furutani,1980Furutani}:
\begin{align}
\tilde{V}_c(r) &\, = \:  \sum_{n=1}^3  \, V_c^n   \,    \left( W_c^n + B_c^n P_{\sigma} - H_c^n  P_{\tau} \right. \nonumber \\[-2.2ex]
& \qquad\qquad\qquad\quad \;\:\;\left.- M_c^n P_{\sigma}P_{\tau} \right)\: e^{-\beta_c^n r^2}  \label{eq.FHT1} \\
\tilde{V}_{\scriptscriptstyle{L\:\!\!S}}(r)  &\, = \: \vec{L}\cdot \vec{S}\;\, \sum_{n=1}^2 \, V_{\scriptscriptstyle{L\:\!\!S}}^n  \, \left( W_{\scriptscriptstyle{L\:\!\!S}}^n - H_{\scriptscriptstyle{L\:\!\!S}}^n  P_{\tau}  \right) \, e^{-\beta_{\scriptscriptstyle{L\:\!\!S}}^n r^2}   \label{eq.FHT2} \\
\tilde{V}_{\scriptscriptstyle{T}}(r) &\, = \:  S_{ij}\:\sum_{n=1}^3 \,  V_{\scriptscriptstyle{T}}^n \, \left( W_{\scriptscriptstyle{T}}^n  - H_{\scriptscriptstyle{T}}^n  P_{\tau} \right) \, r^2 e^{-\beta_{\scriptscriptstyle{T}}^n r^2},   \label{eq.FHT3}
\end{align}
where $r\equiv r_{ij}$ stands for the  distance between the nucleons $i$ and $j$, $\vec{L}$ is the relative orbital angular momentum,  $\vec{S}=(\vec{\sigma}_i+\vec{\sigma}_j) / 2$,  $S_{ij}=3 (\vec{\sigma}_i \cdot \hat{r}) (\vec{\sigma}_j \cdot \hat{r})   - \vec{\sigma}_i \cdot \vec{\sigma}_j$, and $P_{\sigma}$ and  $P_{\tau}$ are spin and isospin exchange operators, respectively. Each part of the interaction is the sum of (two or) three gaussians with different ranges: a short range to account for the  hard core, a long range to mimic the one-pion exchange potential, and an intermediate range. The spin-orbit interaction does not contain a long-range part and is only a sum of two gaussians \cite{1967Tamagaki}. The original parameters of the interaction of Refs. \cite{1979Furutani,1980Furutani} are listed in Table \ref{Table.FHT}; they  were used to reproduce the  binding energy of $^4$He, as well as the nucleon scattering phase shifts of on $A=3,4$ nuclei.

\begin{table}[htb]
\begin{ruledtabular}
\caption{\label{Table.FHT} Parameters of the central, spin-orbit, and tensor interactions of  Ref. \cite{1979Furutani}. The depths $V^n$ are given in MeV for the central and spin-orbit interactions and in MeV\,fm$^{-2}$ for the tensor interaction. The ranges $\beta$ are in fm$^{-2}$. The Wigner, Majorana, Bartlett, and Heisenberg parameters are dimensionless.}
\begin{tabular}{lccccccc }
&  $n$ & $V^n$  & $\beta^n$& $W^n$   & $M^n$  & $B^n$  & $H^n$      \\
	\hline\\[-7pt]
 & 1    & $-$6.0     & 0.160  & $-$0.2363 & 1.1530 & 0.5972 & $-$0.5139                  \\
$V_c$ & 2    & $-$546.0  & 1.127   &  0.4242 & 0.4055 & 0.1404 &  0.030                  \\
 & 3    &  1655.0   & 3.400    &  0.4474 & 0.3985 & 0.1015 &  0.0526             \\
 \hline\\[-7pt]
 \multirow{2}{*}{$V_{\scriptscriptstyle{L\:\!\!S}}$}  & \multicolumn{1}{c}{1} & \multicolumn{1}{c}{1918.0} & \multicolumn{1}{c}{5.0} & \multicolumn{1}{c}{0.5} & \multicolumn{1}{c}{} & \multicolumn{1}{c} {} & \multicolumn{1}{c}{$-$0.5}  \\
                                 & \multicolumn{1}{c}{2}  & \multicolumn{1}{c}{$-$1519.0} & \multicolumn{1}{c}{3.0 } & \multicolumn{1}{c}{0.5} & \multicolumn{1}{c}{} & \multicolumn{1}{c}{}  & \multicolumn{1}{c}{$-$0.5}   \\
\hline\\[-7pt]
 & 1   & $-$16.96      & 0.53     & 0.3277 &&& 0.6723                \\
$V_{\scriptscriptstyle{T}}$ & 2   & $-$369.5   & 1.92       & 0.4102 &&& 0.5898                \\
  &3   &  1688.0    & 8.95      & 0.5   &&& 0.5                   \\
\end{tabular}
\end{ruledtabular}
\end{table}

In order to be applied in the present GSM formalism, the interaction is rewritten in terms of the spin-isospin projectors $\Pi_{ST}$ \cite{1974DeShalit,RingSchuck}:
\begin{align}
V_c(r) &\;  = \;  V_c^{11} \, f_c^{11}(r)  \Pi_{11} \: + \:  V_c^{10} \, f_c^{10}(r) \Pi_{10} \nonumber  \\
& \qquad  + \: V_c^{00} \, f_c^{00}(r) \Pi_{00} \: + \:  V_c^{01} \, f_c^{01}(r) \Pi_{01},  \label{eq.inter1}  \\[1.5ex]
V_{\scriptscriptstyle{L\:\!\!S}}(r) &\;  = (\vec{L}\cdot \vec{S}) \,V_{\scriptscriptstyle{L\:\!\!S}}^{11} \, f_{\scriptscriptstyle{L\:\!\!S}}^{11}(r) \Pi_{11}, \label{eq.inter2}  \\[1.5ex]
V_{\scriptscriptstyle{T}}(r) &\;  = S_{ij} \left[V_{\scriptscriptstyle{T}}^{11}f_{\scriptscriptstyle{T}}^{11}(r) \Pi_{11} + V_{\scriptscriptstyle{T}}^{10} f_{\scriptscriptstyle{T}}^{10}(r) \Pi_{10}\right],\label{eq.inter3}
\end{align}
with the seven interaction strengths in spin-isospin channels,
$V_c^{11}$, $V_c^{10}$, $V_c^{00}$, $V_c^{01}$, $V_{\scriptscriptstyle{L\:\!\!S}}^{11}$, $V_{\scriptscriptstyle{T}}^{11}$, and   $V_{\scriptscriptstyle{T}}^{10}$, remaining  to be optimized. 
The form factors $f^{ST}$  are linear combinations of the original radial form-factors
appearing in Eqs.~(\ref{eq.FHT1}-\ref{eq.FHT3}). They
are normalized to the first parameter $V^1$ for each central, spin-orbit, and tensor terms in order to make them dimensionless. The remaining interaction parameters, such as the gaussian ranges, relative strengths of gaussian components,  and the   Wigner, Majorana, Bartlett, and Heisenberg parameters, have been kept at their original values  as in
Table~\ref{Table.FHT}.

\section{Optimization and uncertainty quantification} \label{Sec.Optimization}

The optimization of the interaction and the assessment of statistical uncertainties were performed according to Ref.~\cite{2014Dobaczewski}. Given a model in which $N_p$ parameters $\vec{p} = \{p_1,.,p_{N_p}\}$ are adjusted to describe $N_d$ observables $\mathcal{O}_i $  $(i=1,.,N_d)$, the optimization procedure is based on the minimization of the penalty function:
\begin{equation}
	\chi^2(\vec{p}) = \sum_{i=1}^{N_d}\left(\frac{\mathcal{O}_i(\vec{p}) - \mathcal{O}_i^{\mbox{\scriptsize{exp}}}}{\delta \mathcal{O}_i} \right) ^2,\label{chiFunction}
\end{equation}  
where $\mathcal{O}_i(\vec{p})$ are  calculated observables and $\mathcal{O} _i^{\mbox{\scriptsize{exp}}}$ are experimental data (fit-observables) used to constrain the model. The adopted errors $\delta \mathcal{O}_i$ include different contributions stemming from  experimental uncertainties,  numerical inaccuracies, and  theoretical errors due to model deficiency. The choice for the theoretical error obviously involves a certain level of arbitrariness even when driven by physical considerations. Part of this arbitrariness can be removed by tuning the adopted errors so that they are consistent with the distribution of the residuals similarly to the case of a purely statistical distribution \cite{2014Dobaczewski}. In particular, one requires the total penalty function to be normalized to the number of degrees of freedom $N_{\mbox{\scriptsize{dof}}}=N_d-N_p$ at the minimum $\vec{p}_0$ \cite{1932Birge}:
\begin{equation}
	\frac{\chi^2(\vec{p}_0)}{N_{\mbox{\scriptsize{dof}}}}  \leftrightarrow 1 . \label{chi2Normalization}
\end{equation}
In the case of a single type of data and on the assumption that experimental and numerical errors are negligible, the condition (\ref{chi2Normalization}) can simply be achieved through a global scaling of the initial adopted errors $\delta \mathcal{O}_i \rightarrow \: \delta \mathcal{O}_i\sqrt{\chi^2(\vec{p}_0) / N_{\mbox{\scriptsize{dof}}}}$. 

With this choice of the normalization condition, one can apply the standard rules of linear regression in statistical analysis and assess quantities such as the covariance matrix and statistical uncertainties. Within the linear regression approximation, the covariance matrix $\mathcal{C}$ can be expressed in terms of the Jacobian~$J$:
\begin{equation}
\mathcal{C} \simeq (J^T J)^{-1}   \;, \quad J_{i\alpha}= \frac{1}{\delta \mathcal{O}_i}\left. \frac{\partial \mathcal{O}_i}{\partial p_\alpha}\right|_{\vec{p}_0},  
\end{equation}
and the covariance between two observables $A$ and $B$ follows as:
\begin{equation}
	\overline{\Delta A \, \Delta B} \simeq  \sum_{\alpha,\beta=1}^{N_p}
                          \left. \frac{\partial A}{\partial p_\alpha} \right|_{\vec{p}_0} \!
                          \mathcal{C}_{\alpha \beta}
                          \left. \frac{\partial B}{\partial p_\beta} \right|_{\vec{p}_0} . \label{covariance}
\end{equation}
In particular, for $A=B$, Eq. (\ref{covariance}) gives the statistical uncertainty on the observable $A$:  $\Delta A = \sqrt{\overline{\Delta A ^2}}$. The dimensionless correlation coefficient \citep{1997Brandt} is defined as:
\begin{equation}
	c_{AB} =  \frac{\overline{\Delta A \, \Delta B}}{\Delta A \,\Delta B}.  \label{correlationCoefficients}
\end{equation}
Applying Eqs. (\ref{covariance}, \ref{correlationCoefficients}) to the model parameters, their statistical uncertainties reduce to $\Delta p_\alpha = \sqrt{ \mathcal{C}_{\alpha \alpha}}$, and the correlation coefficients between two parameters $p_\alpha$ and $p_\beta$ are related to the covariance matrix elements by:
\begin{equation}\label{eq.corr}
  c_{\alpha \beta} = \frac{\mathcal{C}_{\alpha \beta}}
                          {\sqrt{
                                 \mathcal{C}_{\alpha \alpha}
                                 \mathcal{C}_{\beta \beta}}} \:.
\end{equation}
It is worth mentioning that those expressions are valid within the linear regression approximation, which assumes that  the observables $A$ and $B$ behave  linearly with respect to the model parameters around  the minimum $\vec{p}_0$. In principle, Eqs. (\ref{covariance},\ref{correlationCoefficients}) are valid for the model parameters and  the observables to be predicted by the model. However, these do not hold for the postdicted fit-observables as these are  well constrained around the minimum. The uncertainties on the adjusted observables can only be assessed through a full statistical analysis, such as Bayesian inference.

The minimization of the penalty function (\ref{chiFunction}) plays a central part in the optimization process. A good minimization algorithm should be able to cope with two main obstacles: a possible strong intercorrelation between the adjusted observables and the sloppiness \cite{Sloppy}, or indeterminacy, of some parameters, i.e., the fact that some parameters can be weakly constrained by the fit-observables  \cite{Ber05,2008Toivanen,2008Kortelainen,Sto10,2016Nicsic}. For that matter, many minimization methods have been developed recently such as Monte-Carlo algorithms  \cite{2014NavarroPerez,*2014NavarroPerez_Bootstrapping} or the POUNDerS algorithm \cite{Pounders} which has been applied successfully in the contexts of nuclear density functional \cite{UNEDF0,UNEDF1} or chiral interaction \cite {2013Ekstrom} optimizations. In our model, the interactions are linear in strength parameters, and the derivatives can be computed exactly using the Hellmann-Feynman theorem \cite{1939Feynman}. These derivatives can be directly exploited in our optimization algorithm which involves  the Gauss-Newton method combined with the Singular Value Decomposition  (SVD) technique. The Gauss-Newton method is a variation of the standard Newton minimization algorithm to the case of chi-square-like functions. The SVD cures the instability of the Gauss-Newton method which appears when the Jacobian matrix is non-invertible or has a very small determinant, which happens when  fit-observables are highly correlated and/or some parameters are unconstrained. A full description of the method can be found in Ref. \cite{2008Kortelainen}.

\section{Optimization of the core potential}\label{Sec.WS}

\begin{figure}[htb]
  \includegraphics[width=0.9\columnwidth]{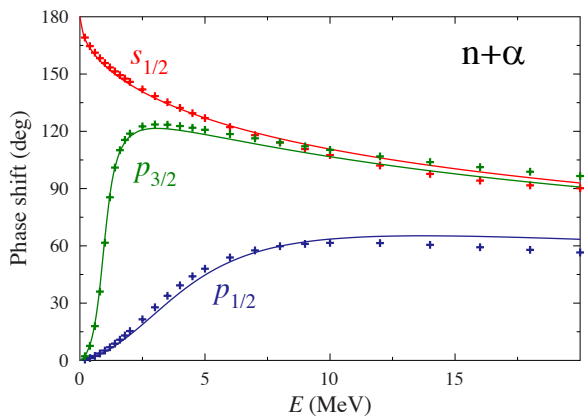}
  \caption{\label{Fig.psn} Optimized $s_{1/2}$ (top, red), $p_{3/2}$ (middle, green) and $p_{1/2}$ (bottom, blue) neutron$-{}^4$He nuclear phase shifts as functions of the energy of the neutron in the laboratory obtained using the Woods-Saxon parameters given in Table \ref{Table.WSparam}. The experimental values represented by crosses are taken from Ref. \cite{1977Bond}.}
\end{figure}

\begin{figure}[htb]
  \includegraphics[width=0.9\columnwidth]{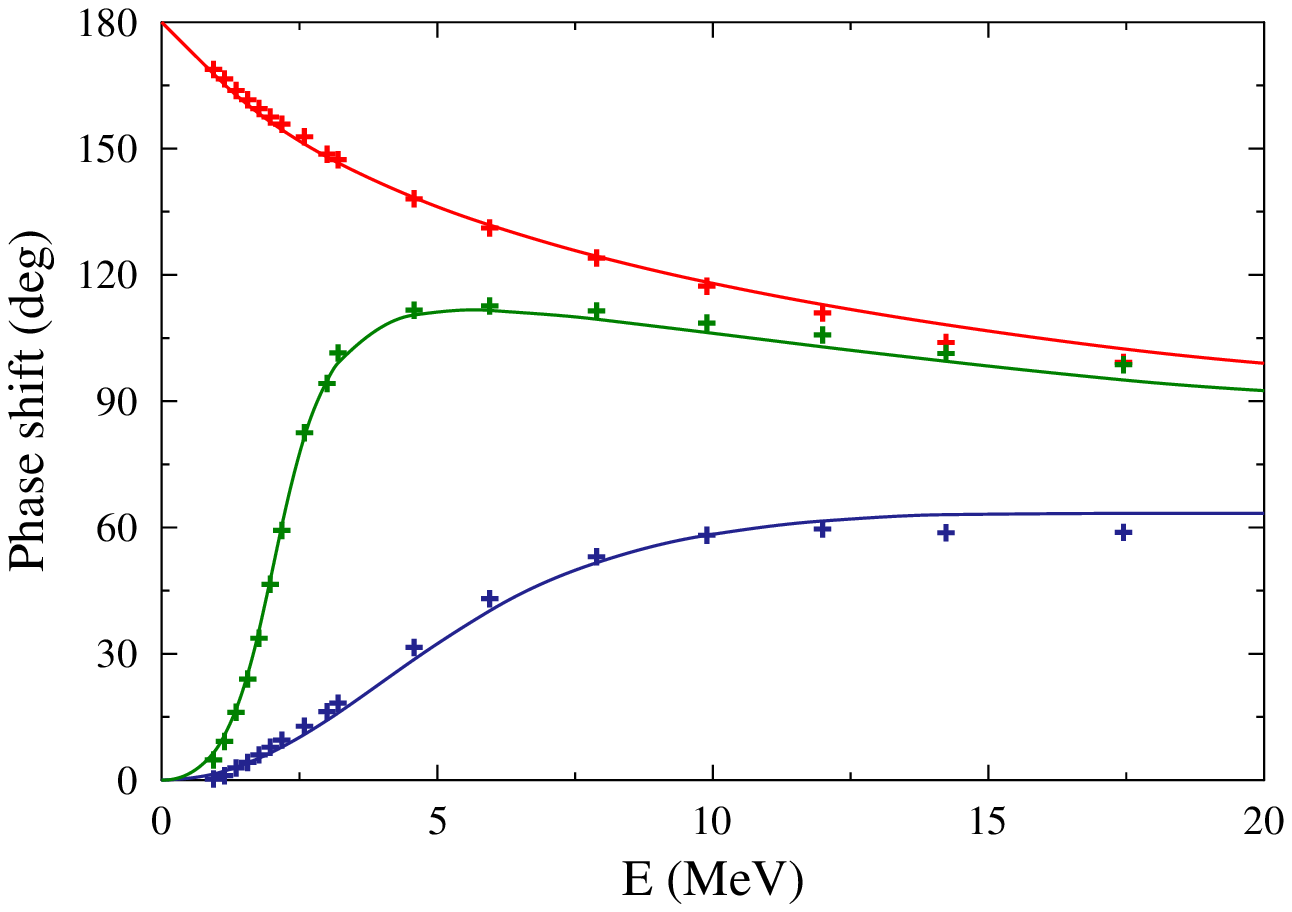}
  \caption{\label{Fig.psp}   Similar as in Fig.~\ref{Fig.psn} but for  the proton-$^4$He scattering. The experimental data from 0 to 3 MeV are taken from Ref.~\cite{1967Brown} and the data from 3 to 20 MeV from Ref.~\cite{1971Schwandt}.}
\end{figure}

The ${}^4$He-nucleon  interaction of Eq.~(\ref{eq.woodsSaxon}) was optimized to the experimental $p_{3/2}$, $p_{1/2}$, and $s_{1/2}$ nucleon-$^4$He scattering phase-shifts up to 20 MeV  \cite{1977Bond,1967Brown,1971Schwandt}. The optimization procedure  yielded a well-converged result
for both protons and neutrons corresponding to a precision of $\Vert \nabla \chi^2\Vert/N_{\mbox{\scriptsize{dof}}} \sim 10^{-12}$. 
To check whether the minimum is global, we repeated the optimization starting from different
parameter sets. The values of the optimized WS parameters and their statistical uncertainties are listed in Table \ref{Table.WSparam}, and the corresponding phase shifts in Figs.~(\ref{Fig.psn},\ref{Fig.psp}).
\begin{table}[htb]
\caption{\label{Table.WSparam}Parameters of the optimized  $^4$He-nucleon  interaction with associated statistical uncertainties. The charge radius $R_{\rm ch}$ was set to the experimental value \cite{Sick} and did not enter the optimization procedure.}
\begin{ruledtabular}
\begin{tabular}{lcc}
Parameter & Neutrons & Protons \\
					\hline\\[-7pt]

$V_0$  (MeV) &  41.9 (10)   & 44.4 (11)\\
$V_{\ell\;\!\!s}$ (MeV\,fm$^2$) &  7.2 (2)  &7.2 (2) \\
$R_0$ (fm) &  2.15 (4) &  2.06 (4) \\
$a$ (fm)  & 0.63 (2) &  0.64 (2) \\
$R_{\rm ch}$ (fm)  &--  & 1.681
\end{tabular}
\end{ruledtabular}
\end{table}
The ${}^4$He core is known to be inert and its optical potential is  well described by a WS model   at low excitation energies \cite{1954Sack}. Not surprisingly, our optimized calculations yield  a very good agreement with the experimental low-energy phase shifts. The small discrepancies seen at $E>15$ MeV can be attributed to the virtual excitations to the excited states of $^4$He which can no longer be neglected at this energy range.

The isospin difference in the parameters of Table~\ref{Table.WSparam}, in particular regarding the WS depth $V_0$, has been well-documented in the literature and is seen experimentally in the 
behavior of the $p_{1/2}$ phase shifts at $E<1.5$\,MeV, where the proton phase shifts remain close to zero (cf. Figs. \ref{Fig.psn} and \ref{Fig.psp}) \cite{1954Sack}. It is to be noted that the GSM approach employs  the COSM coordinates in which the masses that enter the Schr\"{o}dinger equation (\ref{Hamiltonian}) are the reduced masses. This explains
the  deviation from typical values of WS potential parameters, $V_0\simeq 50$\,MeV, found in the literature  \cite{RingSchuck}.  As stated in Section \ref{SSec.GSM}, the calculations performed in COSM coordinates are practically equivalent to those the Jacobi coordinates. This consistency can be illustrated by calculating the energies and widths of the $3/2^{-}$ ground states of $^5$He and  $^5$Li. Table~\ref{Table.spe} demonstrates a good agreement with experimental data, especially given the large widths of the resonant states. 
\begin{table}[htb]
\begin{ruledtabular}
\caption{\label{Table.spe}Energies (in MeV) and widths (in keV) of the $3/2^-$ ground states of $^5$He and $^5$Li calculated using the optimized optical model with parameters listed in Table \ref{Table.WSparam}. The experimental values are taken from Refs.~\cite{2002Tilley,TUNL}.}
\begin{tabular}{lcccc}
 Nucleus & $E$       & $E_{\mbox{\scriptsize{exp}}}$  & $\Gamma$   & $\Gamma_{\mbox{\scriptsize{exp}}}$ \\
 					\hline\\[-7pt]
 $^5$He  & 0.755 & 0.798   & 651 & 648 \\
 $^5$Li  & 1.627 & 1.69   & 1351 & 1230  
\end{tabular}
\end{ruledtabular}
\end{table}

Table~\ref{Table.param.corrs} shows the matrix (\ref{eq.corr}) 
of inter-parameter correlations. Together with uncertainties on parameters given in Table~\ref{Table.WSparam},  this information  can be used in to compute the uncertainties on observables.  
While most parameters are correlated with each other, 
there is a  strong correlation between the WS potential depth and  radius. This is due to the fact that the eigenvalues  of the WS potential are primarily governed  by the volume of its shape and different sets of $(V_0,R_0)$ are quite equivalent. 
\begin{table}[htb]
\begin{ruledtabular}
\caption{\label{Table.param.corrs}Correlation matrix (\ref{eq.corr}) of the optimized  $^4$He-nucleon  interaction for the protons (upper triangular matrix) and neutrons (lower triangular matrix).}
\begin{tabular}{lcccc}
 \diagbox[height=5mm]{$n$}{\raisebox{0.1em}{$p$}}&    \phantom{$-$}$V_0$  & \phantom{$-$}$V_{\ell\;\!\!s}$ & $R_0$  & \phantom{$-$}$a$  \\
	\hline\\[-7pt]
 $V_0$  & \phantom{$-$}1 & \phantom{$-$}0.62   &$-$0.95  & \phantom{$-$}0.59 \\
 $V_{\ell\;\!\!s}$  & \phantom{$-$}0.55 & \phantom{$-$}1  & $-$0.78 & \phantom{$-$}0.81  \\
 $R_0$  & $-$0.95 & $-$0.75  & \phantom{$-$}1 & $-$0.81  \\
 $a$ & \phantom{$-$}0.52 & \phantom{$-$}0.84   & $-$0.75 & \phantom{$-$}1  \\
\end{tabular}
\end{ruledtabular}
\end{table}

\section{Optimization of the two-body interaction}\label{Sec.EffectiveInteraction}

Having defined the core-nucleon potential,  we optimized the two-body interaction (\ref{eq.inter1}-\ref{eq.inter3}) between valence nucleons. The calculations were performed in the $psdf-$configuration space. The $0p_{3/2}$, $0p_{1/2}$ resonant states and the associated  scattering continua were used as the Berggren basis for both protons and neutrons, as well as the  $1s_{1/2}$ and $0d_{5/2}$ resonant states  and the associated continua for  neutrons to account for possible antibound shells and excited states of different parity.

As stated in Sec.~\ref{SSec.GSM},  the basis potential that generates the Berggren basis was adapted for each nucleus. Consequently, the scattering continua were also chosen differently for all nuclei depending on the nature of the $0p_{3/2}$, $0p_{1/2}$, $1s_{1/2}$, and $0d_{5/2}$ Gamow poles. For example, in the cases in which the considered pole and the desired many-body state were  bound, the contour consisted of three segments on the real axis of the momentum plane defined by the points $k_{\mbox{\scriptsize{peak}}}=(0.1,0.0)$ fm$^{-1}$, $k_{\mbox{\scriptsize{mid}}}=(0.2,0.0)$ fm$^{-1}$, and $k_{\mbox{\scriptsize{max}}}=(2.0,0.0)$ fm$^{-1}$. In the case of unbound  s.p. pole, $k_{\mbox{\scriptsize{peak}}}$ and $k_{\mbox{\scriptsize{mid}}}$ were moved  into the complex momentum plane to encompass the resonant state. Finally, in the special case of $^7$He, for which the $0p_{3/2}$ pole is bound but the many-body state is unbound, the corresponding contour was defined by  the points $k_{\mbox{\scriptsize{peak}}}=(0.25,-0.24)$ fm$^{-1}$, $k_{\mbox{\scriptsize{mid}}}=(0.5,0.0)$ fm$^{-1}$, and $k_{\mbox{\scriptsize{max}}}=(2.0,0.0)$ fm$^{-1}$ to generate a many-body configuration space which can describe an unbound many-body state. In all cases, the three segments were discretized with at least 10 Gauss-Legendre points. It is worth noting that the detailed choice of the contour should not influence the results, provided that the key Gamow poles are encompassed by the scattering contour.

The remaining higher-$\ell$ partial waves  can be quite well described using a HO basis \cite{2016Hagen}. For that matter, and to reduce the size of the model space, the remaining $s$, $d$ and $f$ partial waves were spanned by a HO basis with 11 shells ($n^{\rm HO}_{\mbox{\scriptsize{max}}}=10$). In this mixed basis, the natural orbitals were generated as discussed in  Sec.~\ref{SSec.GSM}. The final model space in which the calculations with the natural orbitals were performed, allows at most four particles in the scattering continuum.

The two-body interaction was optimized to the experimental binding energies of the ground states and a few selected excited states of the helium, lithium, and beryllium isotopes shown in Table~\ref{Table.optimizedEnergies}.
\begin{table}[htb]
\begin{ruledtabular}
\caption{\label{Table.optimizedEnergies}Binding energies (relative to  $^4$He; in MeV) and widths (in keV) of the selected states of $A=6-9$ nuclei used in  this work to optimize the two-body GSM interaction.  The experimental values are taken from 
Ref.~\cite{TUNL}. The theoretical values were obtained  
using the  interaction parameters of Table \ref{Table.InterParam}.  Note that the widths of the listed unbound states did not enter the optimization procedure, i.e., those represent genuine predictions.}
\begin{tabular}{cccccc}
 Nucleus & State & $E$       & $E_{\mbox{\scriptsize{exp}}}$  & $\Gamma$   & $\Gamma_{\mbox{\scriptsize{exp}}}$ \\
	\hline\\[-7pt]
 $^6$He & $0^+$ & $-$1.063 & $-$0.973  && \\
 $^6$He & $2^+$ & \phantom{$-$}0.938 & \phantom{$-$}0.824   & 168 & 113(20) \\
 $^7$He & $3/2^-$ & $-$0.578 & $-$0.528  & 178 & 150(20) \\
 $^8$He & $0^+$ & $-$3.225 & $-$3.112 &&\\
	\hline\\[-7pt]
 $^6$Li & $1^+$ & $-$3.724 & $-$3.699  && \\
 $^6$Li & $0^+$ & $-$0.054 & $-$0.136 &  &  \\
 $^7$Li & $3/2^-$ & $-$10.688 & $-$10.949 & &\\
 $^7$Li & $1/2^-$ & $-$10.359 & $-$10.471 & &\\
 $^8$Li & $2^+$ & $-$13.350 & $-$12.982 & & \\
 $^9$Li & $3/2^-$ & $-$16.677 & $-$17.046 & &\\
	\hline\\[-7pt]
 $^6$Be & $0^+$ & \phantom{$-$}1.390 & \phantom{$-$}1.371 & 21 & 92(6)\\
 $^7$Be & $3/2^-$ & $-$8.977 & $-$9.305 & &\\
 $^8$Be & $0^+$ & $-$28.572 & $-$28.204 & 0  & 0.0056(3)\\
 $^9$Be & $3/2^-$ & $-$30.230 & $-$29.870 & & \\
 $^9$Be & $1/2^+$ & $-$27.747 & $-$28.186 & 0 & 217(10) 
\end{tabular}
\end{ruledtabular}
\end{table}
The binding energies span a large range from approximately $-30$ MeV to $+2$ MeV, and different types of states are involved: bound states, resonances, and  halo states (ground state of $^6$He). The $1/2^+$ state of $^9$Be has been chosen to probe the $s_{1/2}$ shell. 

The optimization yielded a $\chi^2$ minimum  with a precision $\Vert \nabla \chi^2\Vert/N_{\mbox{\scriptsize{dof}}} \sim 10^{-4}$   limited by the SVD cutoff value (see below). We restarted the optimization using 
different points in the parameter space to assure that the robust solution has been found. The optimized interaction parameters are listed  
in Table~\ref{Table.InterParam} together with the associated uncertainties. 
\begin{table}[htb]
\begin{ruledtabular}
\caption{\label{Table.InterParam}Optimized parameters of the two-body interaction (\ref{eq.inter1}-\ref{eq.inter3}) together with  their statistical uncertainties.}
\begin{tabular}{@{\hspace{3em}}lc@{\hspace{3em}}}
Parameter & Value \\
	\hline\\[-7pt]
$V_c^{11}$ (MeV) &  $-$3.2 (220) \\
$V_c^{10}$  (MeV) &  $-$5.1  (10)  \\
$V_c^{00}$ (MeV) &  $-$21.3 (66) \\
$V_c^{01}$ (MeV) &  $-$5.6 (5) \\
$V_{\scriptscriptstyle{L\:\!\!S}}^{11}$ (MeV) & $-$540 (1240) \\
$V_{\scriptscriptstyle{T}}^{11}$ (MeV\,fm$^{-2}$) &    $-$12.1 (795) \\
$V_{T}^{10}$  (MeV\,fm$^{-2}$) &  $-$14.2 (71)
\end{tabular}
\end{ruledtabular}
\end{table}

As some parameters are weakly constrained by the current dataset, the SVD procedure played an important part in our optimization.
To account for their different units and orders of magnitude,  the parameters were normalized to the value of one during the SVD procedure, that is $p_{\alpha}\rightarrow \tilde{p}_{\alpha}= 1$, $J_{i\alpha} \rightarrow \tilde{J}_{i\alpha} = p_{\alpha}J_{i\alpha}$. 
Table \ref{Table.SVD} lists the singular values (square roots of the eigenvalues of the Hessian matrix $\tilde{J}^T\tilde{J}$) at the minimum together with the corresponding eigenvectors. 
\begin{table}[htb]
\begin{ruledtabular}
\caption{\label{Table.SVD}Singular values $s_n$ and the corresponding eigenvectors of the normalized Hessian matrix $\tilde{J}^T\tilde{J}$ with respect to the parameters at the minimum.  The main components are written in boldface. The SVD cutoff  separates the relevant space generated by the eigenvalues 1-4 from the irrelevant space. The displayed values are computed at the $\chi^2$ minimum but exhibit similar pattern during the optimization procedure.}
\begin{tabular}{cc|ccccccc}
$n$ & $s_n$ & \phantom{$-$}$V_c^{11}$ & \phantom{$-$}$V_c^{10}$  & \phantom{$-$}$V_c^{00}$& \phantom{$-$}$V_c^{01}$ & \phantom{$-$}$V_{\scriptscriptstyle{L\:\!\!S}}^{11}$  & \phantom{$-$}$V_{\scriptscriptstyle{T}}^{11}$ &  \phantom{$-$}$V_{T}^{10}$      \\
	\hline\\[-7pt]
1 & 243 & \phantom{$-$}0.00 & \phantom{$-$}\textbf{0.82} & $-$0.03 & \phantom{$-$}\textbf{0.53} & \phantom{$-$}0.00 &\phantom{$-$}0.00 & \phantom{$-$}0.23        \\
2 & 43.0 & \phantom{$-$}0.00 & \textbf{$-$0.49} & $-$0.02 &\phantom{$-$}\textbf{0.85} & \phantom{$-$}0.00 & $-$0.01 & $-$0.19       \\
3 & 7.06 & $-$0.04 & $-$0.16 & \phantom{$-$}\textbf{0.79} & \phantom{$-$}0.05 & \phantom{$-$}0.04 & $-$0.07 & \phantom{$-$}\textbf{0.58}        \\
4 & 3.94 & \phantom{$-$}0.02 & $-$0.25 & \textbf{$-$0.61} & \phantom{$-$}0.01 & $-$0.09 & $-$0.04 & \phantom{$-$}\textbf{0.75}        \\
\hline\\[-7pt]
5 & 0.57 & $-$0.23 & $-$0.02  & $-$0.09 & \phantom{$-$}0.00  & \phantom{$-$}\textbf{0.97} & $-$0.01 & \phantom{$-$}0.04        \\
6 & 0.20 & \phantom{$-$}\textbf{0.65} & $-$0.03 &\phantom{$-$}0.04 & \phantom{$-$}0.01 & \phantom{$-$}0.16 &\phantom{$-$}\textbf{0.74} & \phantom{$-$}0.06        \\
7 & 0.12 & \phantom{$-$}\textbf{0.73} & \phantom{$-$}0.01 &\phantom{$-$}0.00 & \phantom{$-$}0.00 & \phantom{$-$}0.16 &\textbf{$-$0.66} & $-$0.04    
 \end{tabular}
\end{ruledtabular}
\end{table}

The eigenvectors associated with large singular values define the  directions along which the penalty function exhibits the largest variations. Following SVD, the parameter space is reduced to a smaller (relevant)  space defined by the singular values greater than a given cutoff value $s_{\rm min}$. In the case considered, a large value of $s_{\rm min}=1$ was needed for the optimization procedure to converge, reducing the parameter space to four main directions. Table~\ref{Table.SVD} also shows that the two central-potential parameters $V_c^{10}$ and $V_c^{01}$  are the two parameters which primarily govern the optimization, as well as $V_c^{00}$  $V_{T}^{10}$ to a lesser extent. The three $(ST)=(11)$ parameters are poorly constrained by the experimental dataset chosen.  More experimental data of different kinds, such as charge and matter radii and electromagnetic moments, will be useful in the future developments  to constrain these parameters. At this point, the freedom on the sloppy parameters can be utilized  to fine-tune the interaction to reproduce experimental reaction thresholds. 

\begin{figure}[htb]
  \includegraphics[width=0.8\columnwidth]{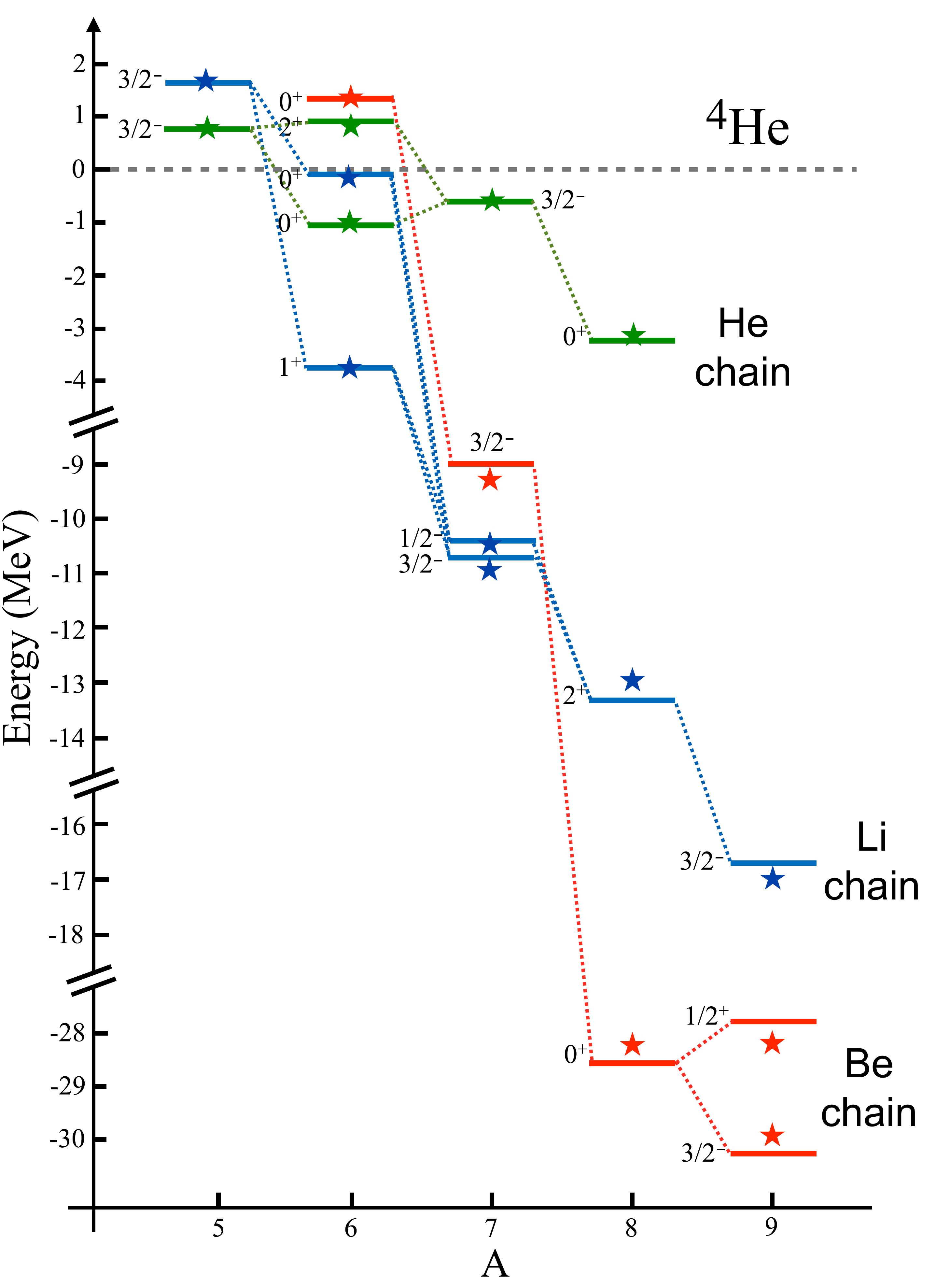}
  \caption{\label{Fig.OptimizedEnergies}  Energies of the helium, lithium and beryllium isotopic chains with $A\le 9$   computed using the optimized GSM interaction with parameters given in Tables \ref{Table.WSparam} and  \ref{Table.InterParam}. The experimental values shown by stars are taken from Ref. \cite{TUNL}. The widths of unbound states  are listed in Table \ref{Table.optimizedEnergies}.}
\end{figure}
The results of the optimization are shown in Fig.~\ref{Fig.OptimizedEnergies} and listed in Table~\ref{Table.optimizedEnergies}. Overall, the quality of the optimization is excellent,  with a root mean square deviation (rms) of 250 keV. The helium chain, whose energetics  depends almost exclusively on a single  parameter $V_c^{01}$, is well described with a rms deviation of 95 keV. The $T=0$ nuclear interaction is responsible for clusterization effects and probably demands the inclusion of higher partial waves than $\ell=3$ for a better description. This explains why the optimization slightly deteriorates for Li and Be isotopes.  In any case, an overall agreement with experiment over such a large range of energies is quite satisfactory and makes this interaction an excellent starting point for detailed structural and reaction studies.
It is also worth noting that, even if they do not enter the set of fit-observables, the widths of the unbound states are  described fairly well in spite of the fact that they  are extremely dependent on the threshold energies. 

The correlation coefficients (\ref{eq.corr}) for the two-body interaction parameters are listed in Table \ref{Table.CorrelationsFHT}. This table can be used to  obtain the associate covariance matrix needed  to assess the uncertainties on predicted observables. 
The two main interaction parameters $V_c^{10}$ and $V_c^{01}$ are strongly anticorrelated. The values that are related to the sloppy  parameters should not be taken too rigorously as they are computed within the linear regression framework. Only a fully-consistent statistical study, based, e.g., on Bayesian techniques,  can fully assess  correlations related to these parameters. 
\begin{table}[htb]
\begin{ruledtabular}
\caption{\label{Table.CorrelationsFHT}Correlation coefficients between the two-body interaction parameters.}
\begin{tabular}{  l|ccccccc }
 & \phantom{$-$}$V_c^{11}$ & \phantom{$-$}$V_c^{10}$  & \phantom{$-$}$V_c^{00}$& \phantom{$-$}$V_c^{01}$ & \phantom{$-$}$V_{\scriptscriptstyle{L\:\!\!S}}^{11}$  & \phantom{$-$}$V_{\scriptscriptstyle{T}}^{11}$ &  \phantom{$-$}$V_{T}^{10}$      \\
\hline\\[-7pt]
$V_c^{11}$ & \phantom{$-$}1 &\phantom{$-$}0.24&\phantom{$-$}0.26&$-$0.44&\phantom{$-$}0.63& $-$0.45& $-$0.25\\
$V_c^{10}$ & \phantom{$-$}0.24 & \phantom{$-$}1 &$-$0.22&$-$0.92&\phantom{$-$}0.01&$-$0.89&$-$0.99 \\
$V_c^{00}$ &  \phantom{$-$}0.26 &$-$0.22& \phantom{$-$}1 &\phantom{$-$}0.30& $-$0.21&\phantom{$-$}0.38&\phantom{$-$}0.21\\
$V_c^{01}$ & $-$0.44 & $-$0.92 & \phantom{$-$}0.30 & \phantom{$-$}1 & $-$0.17& \phantom{$-$}0.96&\phantom{$-$}0.89\\ 
$V_{\scriptscriptstyle{L\:\!\!S}}^{11}$ &\phantom{$-$}0.63 & \phantom{$-$}0.01 & $-$0.21 & $-$0.17 & \phantom{$-$}1 &$-$0.28  & $-$0.04   \\
$V_{\scriptscriptstyle{T}}^{11}$ & $-$0.45 & $-$0.89 &\phantom{$-$}0.38 & \phantom{$-$}0.96 & $-$0.28 & \phantom{$-$}1 &\phantom{$-$}0.88 \\
  $V_{T}^{10}$ & $-$0.25 & $-$0.99 & \phantom{$-$}0.21 & \phantom{$-$}0.89 & $-$0.04 & \phantom{$-$}0.88 & \phantom{$-$}1 
\end{tabular}
\end{ruledtabular}
\end{table}

\section{First applications} \label{Sec.Applications}

The optimized interaction presented in the previous sections sets the path for a variety of structure and reaction applications across the $A\simeq 5-12$ nuclei. Such applications will be presented in forthcoming studies. In this section, we present representative  applications of the optimized interaction to different structural properties.

\subsection{Two-nucleon correlation densities in $^6$He and $^6$Li}
\label{SSec.Correlation_densities}

Pairing correlations are very important in nuclei close to the neutron drip line as they can give rise to a significant stabilization of weakly bound nuclei through the continuum coupling \cite{Bel87,Doba96,Fay00,grasso01_270,Rot09a}.
Two-nucleon  correlations can be evaluated through the  correlation density 
\cite{1991Bertsch,2005Hagino,2011Papadimitriou,2017Simin}
$\rho_{\scriptscriptstyle{N\:\!\!N}}(r,\theta) = \langle \Psi \vert \delta(r_1-r)\delta(r_2-r)\delta(\theta_{12}-\theta)\vert \Psi\rangle$, in which $r_1$ and $r_2$ are the positions of the first and second nucleon respectively and $\theta_{12}$ the opening angle between the two nucleons. 
Here, we follow the normalization convention of Ref.~\cite{2011Papadimitriou} in which the
Jacobian $8\pi^2 r^2 r'^2 \sin\theta$ is incorporated into the definition of $\rho_{\scriptscriptstyle{N\:\!\!N}}$, i.e., it does not appear explicitly.

\begin{figure}[htb]
  \includegraphics[width=\columnwidth]{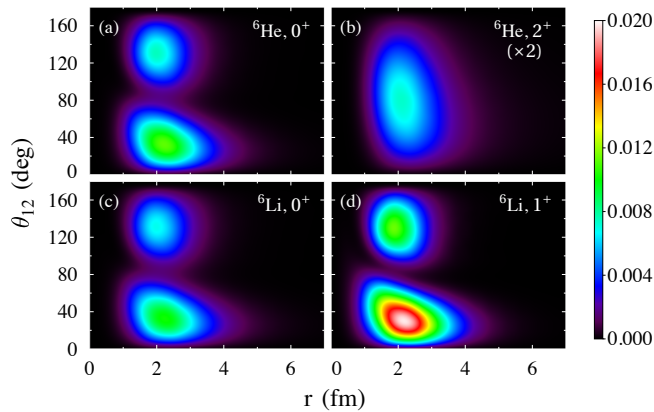}
  \caption{\label{Fig.CorrelationDensities}  Two-nucleon correlation densities (in fm$^{-2}$) calculated for states in  $^{6}$He and $^6$Li using the optimized GSM interaction. For the $2^+$ state in $^6$He  the density was multiplied by a factor of two to maintain the same scale as in other panels.}
\end{figure}

Figure~\ref{Fig.CorrelationDensities} shows the calculated pair correlation densities for the states of   $^{6}$He and $^6$Li  that entered the interaction optimization.
As expected, the  analog $0^+$  states shown in panels (a) and (c) are predicted to have similar correlation densities. Our results are  in full agreement with the conclusions of Refs.~\cite{2005Hagino,2007Horuichi,2011Papadimitriou,2017Simin} pertaining
to the coexistence of  di-nucleon and cigar-like configurations (at small and large opening angles, respectively) and  the large radial extension. Such behavior  is absent for the $2^+$ resonance of $^6$He  shown in panel (b),  for which the valence neutrons are predicted  to be weakly correlated \cite{2011Papadimitriou}.  As seen in Fig.~\ref{Fig.CorrelationDensities}(d), the  strong $T=0$ interaction in the $1^+$  g.s. of $^6$Li gives rise to a deuteron-like structure 
\cite{2007Horuichi,2017Simin}. This result concurs with studies that describe  this state as a deuteron orbiting in the potential generated by the alpha core \cite{1961Kopaleishvili,1969Truoel}.

\subsection{Excited states with uncertainty quantification} \label{SSec.Excited_states}

\begin{figure}[htb]
  \includegraphics[width=\columnwidth]{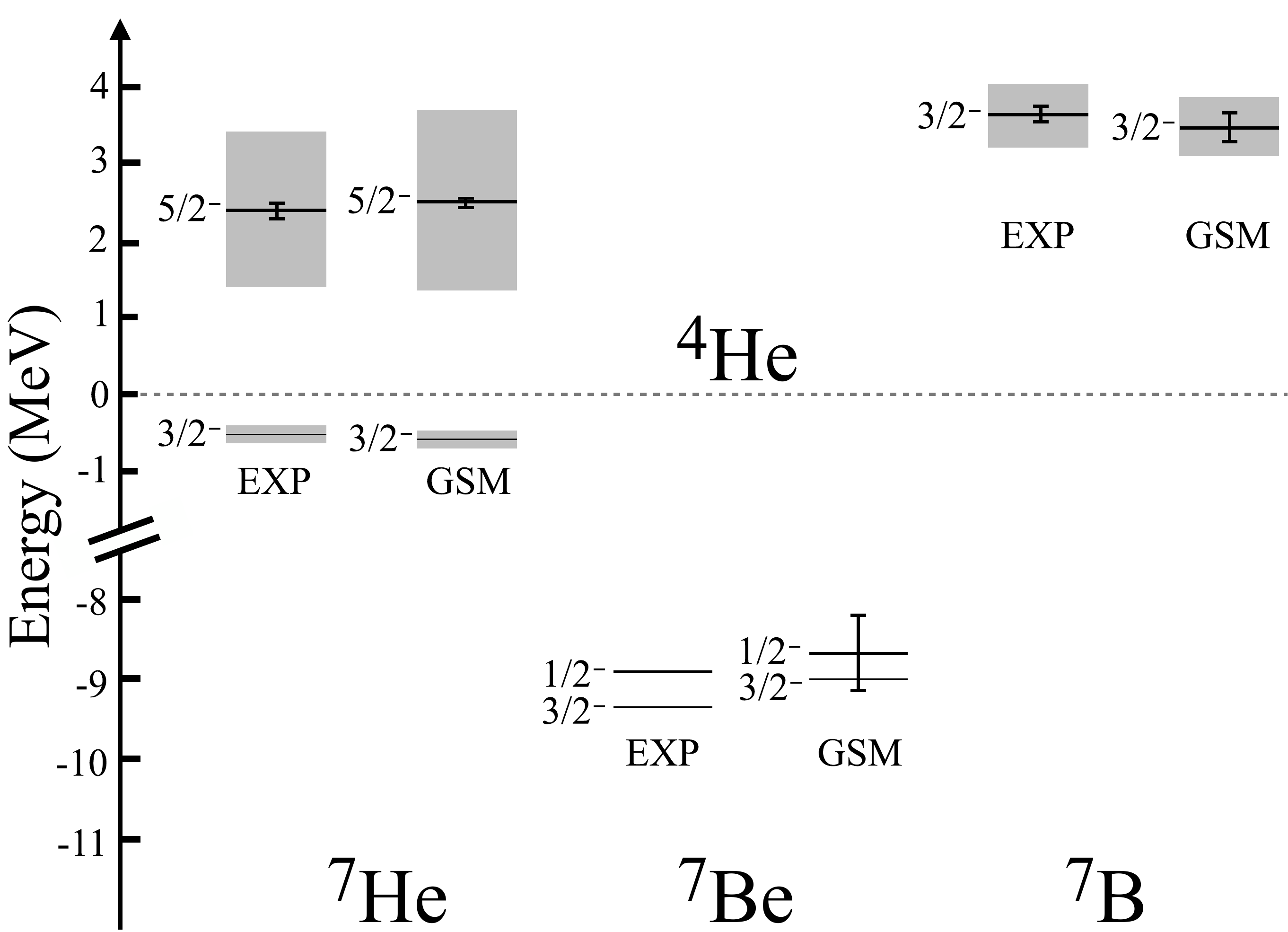}
  \caption{Spectra of $^7$He, $^7$Be,  and $^7$B. The experimental values are taken from  Ref.~\cite{TUNL}. The  g.s. energies  of $^7$He and $^7$Be were included in the optimization, hence are shown without uncertainties. The uncertainties on the widths, not shown in the figure, are listed in Table \ref{Table.applications}.  \label{Fig.A7}}
\end{figure}

\begin{figure}[htb]
  \includegraphics[width=\columnwidth]{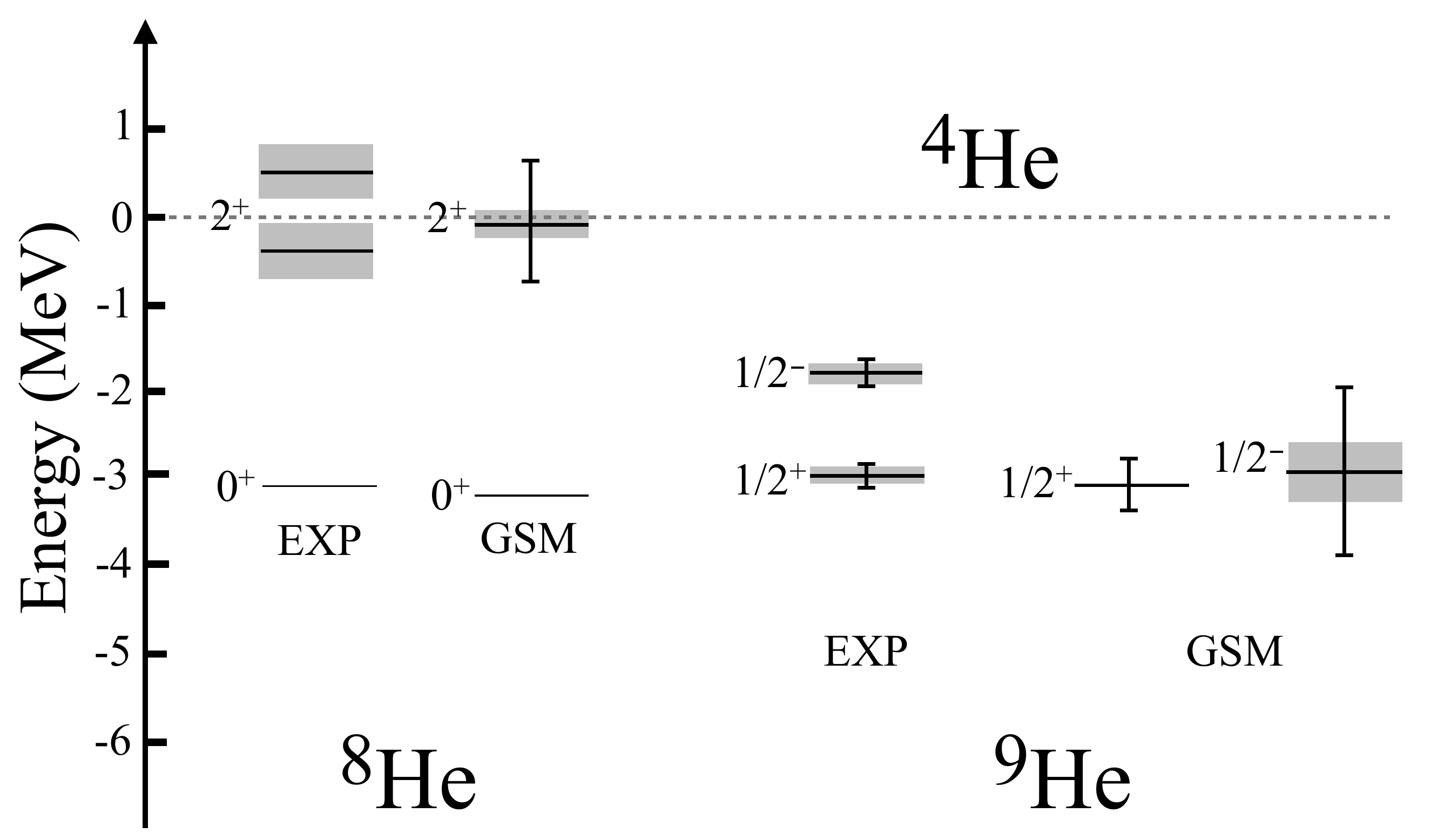}
  \caption{Spectra of $^8$He and $^9$He. The experimental values come from
  \cite{TUNL} ($^8$He) and  \cite{2013Kalanee} ($^9$He).  The energy of the $2^+$ of $^8$He is unresolved experimentally and the two presently adopted values are shown. The ground state of $^8$He was included in the optimization and is shown without uncertainties.  The uncertainties on the widths, not shown in the figure, are listed in Table \ref{Table.applications}. \label{Fig.He} }
\end{figure}

In this section, we  present  predictions of
the optimized GSM potential to 
several  excited  states of light nuclei: the first excited states of $^7$He and $^7$Be and the ground state of $^7$B (Fig. \ref{Fig.A7}),  and the spectra of $^{8}$He and $^9$He (Fig. \ref{Fig.He}). The  energy and widths  are listed in Table \ref{Table.applications} together with their uncertainties.

\begin{table}[htb]
\begin{ruledtabular}
\caption{\label{Table.applications} Calculated values of  $A=7$ nuclei and the Helium nuclei. The experimental values come from \cite{TUNL,2013Kalanee}. All energies are given with respect to the $^4$He core. The energies are given in MeV, the widths in keV. The theoretical and experimental uncertainties on the widths have different meanings and should not be compared (see text).}
\begin{tabular}{lcccc}
 State  & $E$    & $E_{\mbox{\scriptsize{exp}}}$  &  $\Gamma$ & $\Gamma_{\mbox{\scriptsize{exp}}}$ \\
\hline\\[-7pt]
 $^7$He, $5/2^-$ & $+$2.50 (2)  & $+$2.39 (9)  & 2250 (280) & 1990 (170) \\
 $^7$Be, $1/2^-$ & $-$8.67 (45)  & $-$8.88  & &   \\
 $^7$B, $3/2^-$ & $+$3.42 (21)  & $+$3.58 (7)  & 740 (450) & 801 (20) \\
 $^8$He, $2^+$ & $-$0.10 (75) &$-$0.41 / $+$0.49    & 290 (1010) & 600 (200) \\
 $^9$He, $1/2^+$ & $-$3.12 (31) & $-$2.93 (9)   & $\sim$0   &  180 (160)\\
 $^9$He, $1/2^-$ & $-$2.98 (102) &$-$1.88 (12)    & 630 (330)  & 130 (170)  
\end{tabular}
\end{ruledtabular}
\end{table}

Theoretical uncertainties have been calculated  using  the covariance matrices for the one-body (core-nucleon, $N$) and two-body ($NN$) potentials, and they can be expressed as
$\Delta E =\sqrt{\Delta E_N^2 + \Delta E_{NN}^2}$. For all $A<8$ states, the contribution to the uncertainty $\Delta E$ coming from the core-nucleon  part is quite negligible, $\Delta E_{N} < 0.07$\, MeV. This is due to the fact that those states primarily involve  the $p_{3/2}$ bound/resonant shell, which is  well constrained by the core-nucleon potential  optimization. The $p_{1/2}$ and $s_{1/2}$ pole shells are less constrained and result in  larger values of $\Delta E_N$ for the  $1/2^+$ and  $1/2^-$ states of $^9$He (0.13 MeV and 0.59 MeV, respectively). However,  the major part of the uncertainty always comes from the two-body potential.

Both the  energy and width of the first excited $5/2^-$ state of $^7$He are  well reproduced by the GSM interaction, with a small uncertainty. It is important to note that the uncertainties on calculated  widths should be understood as the range of  values when the parameters move within the interval of confidence \cite{2014Dobaczewski}. 
Since the  widths increase very quickly with energy when the state is unbound, the related uncertainties are usually large.
Consequently, those values should not be directly compared to  experimental uncertainties. The $3/2^-$ ground state of $^7$B is also well reproduced as expected, as its mirror state (the ground state of $^7$He) was included in the optimization. Finally, the agreement of the  $1/2^-$of $^7$Be with the experimental value is also  good within the (quite large) uncertainty of 
450\,keV. This significant uncertainty comes from the fact that the less-constrained parameters comes into play. 

The comprehensive discussion  of the helium chain will be the subject of a future study but it is clear from Fig.~\ref{Fig.He} and Table \ref{Table.applications} that the optimized GSM interaction does well in this case. The energy of the $2^+$ of $^8$He is unresolved experimentally with the  two values $-0.41$ MeV and $+0.49$ MeV presently adopted \cite{TUNL}. Our prediction is consistent  with both, due to the  large statistical uncertainty of 750\,keV.  The nucleus $^9$He is a  difficult system  to be studied experimentally. Indeed, as discussed in Ref.~\cite{2013Kalanee}  results of various experiments contradict  each other. As our aim is not to make a thorough analysis of this complex nucleus, we took experimental data of Ref. \cite{2013Kalanee} to compare with our predictions. The uncertainties on predicted values are quite large, in particular for the $1/2^-$ state, as it is the first time that the $p_{1/2}$ resonance shell is occupied in our calculations. This also explains the small uncertainty on the $1/2^+$ state as the $1/2^+$ state of $^9$Be was included in the optimization.  Base on the  the mean values, our interaction  predicts the ground state as a $1/2^+$. This state is calculated too close to the neutron-emission threshold to acquire a significant width.  In any case, this shell inversion predicted by the optimized interaction   is a very interesting feature for future studies of the heavier Li and Be nuclei.

\section{Conclusions and outlook} \label{Sec.conclusions}

In this paper, we have derived a  Gamow Shell Model interaction  to describe very light nuclei in the $psdf$ valence space. The one-body part of the potential reproduces well nucleon-$^4$He  phase shifts, as well as separation energies of $^5$He and $^5$Li.
The two-body interaction contains  central, spin-orbit, and tensor terms; its seven parameters  were optimized to  selected experimental energies of $6\le A \le 9$ nuclei. Based on SVD analysis, we conclude that only four interaction parameters are reasonably constrained by the binding energies. The remaining three parameters, representing $(S=1, T=1)$ strengths, are sloppy; hence new data, different from binding energies, are needed to limit them.

Overall,  a very satisfactory rms deviation from experiment of 250 keV was obtained. The energies of helium isotopes depends almost exclusively on a single  parameter $V_c^{01}$; here, the rms deviation is 95 keV.  This explains why schematic contact forces used in our previous studies
\cite{2002MichelPRL,MICHEL2005} were so successful in explaining binding energies of the He chain.
In future GSM studies of reactions and decays, the relative freedom on sloppy parameters can be utilize to  fine-tune them  to reproduce experimental reaction thresholds. 
Such a strategy, based on controlled GSM extrapolations, has been employed in the recent GSM
applications to drip line nuclei  \cite{FossezMg,FossezOxygen}.
Considering the interaction optimized in this work as a starting point, its predictive power is already  satisfactory as judged by results for two-nucleon densities and energies $A=7$ isotopes and
$^{8,9}$He presented in Sec.~\ref{Sec.Applications}.

An important feature of the new interaction is the associated uncertainty quantification. The covariance matrices computed at $\chi^2$ minimum allow for error estimates on  calculated observables. By analyzing theoretical errors, new insights can be obtained. A good case in point is
the  energy of the $2_1^+$ state in $^8$He.  While the mean value of the calculated energy tends to favor  one experimental result, the range of uncertainties does not allow to rule out another experimental outcome. We also see that some states, such as  the $1/2_1^-$ level in  $^9$He are poorly constrained and this calls for better constraints on  $(S=1, T=1)$ interaction strengths. 

The new GSM interaction sets the path for future investigations  of key dripline nuclei, including structural studies of  halo structures, dineutron and quasi-deuteron configurations, antibound states, as well of reaction studies, including radiative capture reactions within  the GSM-coupled channel formalism \cite{2014Jaganathen,2015Fossez,2017Dong}. In parallel, we intend to carry out a full Bayesian analysis of the GSM interaction, including new kinds of fit-observables. In this way, we hope to improve the fidelity of our model.

\begin{acknowledgments}
We are grateful to Jason Sarich and Stefan Wild for their insights pertaining to numerical optimization. This work was supported by the U.S.\ Department of Energy, Office of
Science, Office of Nuclear Physics under award numbers 
DE-SC0013365 (Michigan State University),  DE-SC0008511 (NUCLEI SciDAC-3 collaboration), and DE-FG02-10ER41700 (French-U.S. Theory Institute for Physics with Exotic Nuclei).
The computations were performed on the Rhea and EOS computer machines at the Oak Ridge Leadership Computing Facility at the Oak Ridge National Laboratory which are supported by the Office of Science of the U.S. Department of Energy. 
\end{acknowledgments}

\bibliography{Effective_interaction}

\end{document}